# XLSearch: A Search Engine for Spreadsheets

Michael Kohlhase, Corneliu Prodescu
Jacobs University Bremen
http://kwarc.info

Christian Liguda DFKI Bremen
http://www.dfki.de/cps/staff/liguda

**ABSTRACT**

*Spreadsheets are end-user programs and domain models that are heavily employed in administration, financial forecasting, education, and science be- cause of their intuitive, flexible, and direct approach to computation. As a result, institutions are swamped by millions of spreadsheets that are becoming increasingly difficult to manage, access, and control.*

*This note presents the XLSearch system, a novel search engine for spread- sheets. It indexes spreadsheet formulae and efficiently answers formula queries via unification (a complex query language that allows metavariables in both the query as well as the index). But a web-based search engine is only one application of the underlying technology: Spreadsheet formula export to web standards like MathML combined with formula indexing can be used to find similar spreadsheets or common formula errors.*

## 1. INTRODUCTION

Spreadsheets are end-user programs and domain models that are heavily employed in administration, financial forecasting, education, and science because of their in- tuitive, flexible, and direct approach to computation. It has been estimated that each year tens of millions of professionals and managers create hundreds of millions of spreadsheets [Pan00]. But we have hardly any tools to mine this immense body of reified knowledge, models, and programmatic experience.

Existing tools center around risk management for spreadsheets via **spreadsheet audits** that create spreadsheet inventories for an organization, estimate risks of in- dividual spreadsheets, and introduce best practices for risk control (see e.g. [Bur08; NO01]), **code reviews** that semi-automatically detect risky parts and practices in spreadsheets and try to ameliorate them, and **test methodologies** that semi- automatically generate test cases for spreadsheets, see e.g. [Rot+01]. Except for the first step in spreadsheet audits, all of these tools are *local* – i.e. apply to single spreadsheets. A notable exception is the EUSES spreadsheet corpus and the statistics gathered for it in [FR05]. However, even this corpus only contains ca. 4.500 spreadsheets, a number which is multiple orders of magnitude smaller than the spreadsheet inventories of large organizations or what is known to search engines: A spreadsheet auditor reported $10^7$ spreadsheets in a single fortune-50 company at EuSpRIG 2010 and a Google search for filetype:xls reports $1.5 \times 10^7$ hits.

For *global services on spreadsheets* we need tools that scale to very large corpora. In practice, this means two things: *i*) standardized, web-scalable representation formats and *ii*) sub-linear processing algorithms. In this paper, we provide both for the case of spreadsheet formulae, and apply this to a concrete application: the XLSearch engine, which allows to efficiently find spreadsheets by querying for their formulae.

**Organization** In the next section, we will present a machine-understandable vocabulary for the ca. 360 functions, constants, and references used in current spread-sheet programs; this acts as the basis for representing spreadsheet formulae as con- tent MathML expressions. This representation allows us to utilize a pre-existing retrieval engine for mathematical formulae (the MathWebSearch system), which we will describe in section 3 to make this paper self-contained. Section 4 presents an application that harvests formulae and result fragments from a spreadsheet for indexing in MathWebSearch. In Section 5, we describe the XLSearch system, a novel search engine for spreadsheets as one possible application we can build with these components. Section 6 concludes the paper and discusses other applications of the combination of MathML representations and indexing/querying.

**Running Example** To make the technical exposition more coherent, we will use the following situation as a running example:

*Semantex Inc*, a successful financial consulting company has just changed its financial forecasting policy from linear extrapolation to second-order La- grange extrapolation and is now faced with changing the spreadsheets it is using for forecasting. This change impacts everything from the reporting spreadsheets to tables embedded into powerpoint presentations. Fortunately, *Semantex Inc* has recently carried out a spreadsheet audit and thus has a good overview over all documents that contain spreadsheet tables.

In such a situation, a spreadsheet formula search engine like XLSearch can help, since it can search for variants of the linear extrapolation formula

$$f(x) \sim f(a) + \frac{x-a}{b-a}(f(b) - f(a))$$

for a function $f$ from its values at $a \leq b$. Even though this example was chosen more for expository qualities than for business realism, it already reveals many qualities of the solution.

2. **SPREADSHEET FORMULAE IN MATHML: SPSHP ONTOLOGY**

MathML [Aus+10] is a W3C standard for the representation of mathematical for- mulae. It contains two sub-languages: *i*) "presentation MathML" for the layout of mathematical formulae – this supports the high-quality presentation of mathematical formulae in browsers and XML-based publishing workflows, and *ii*) "content MathML" for the representation of the functional structure of formulae – this sup- ports interoperability between mathematical software systems. The latter is relevant for our purposes in this paper. Content MathML represents formulae as operator trees consisting of applications of functions to arguments (the apply elements in Figure 1), variables, numbers (mn elements), strings, and symbols. The latter are represented by csymbol elements; the meaning of a symbol is specified by referencing a content dictionary (CD), which provides information about properties of the functions, definitions, notation definitions and types, identifying the concept in the CD by name (the text content of the csymbol element).

```
<math   xmlns="http://www.w3.org/1998/Math/MathML"
    cdgroup="http://oaff.info/spshp/">
  <apply>
    <csymbol   cd="spsht−arith">times</csymbol>
    <apply>
      <csymbol   cd="spsht−arith">sum</csymbol>
      <apply>
        <csymbol  cd="spshform">range</csymbol>
        <mn>1</mn><mn>5</mn><mn>1</mn><mn>8</mn>
      </apply>
    </apply>
    <mn>2</mn>
  </apply>
</math>
```

**Figure 1**: SUM(A5:A8)*2 in content MathML

So the main task in defining a MathML representation for spreadsheet formulae lies in providing a set of CDs that specify the underlying vocabulary.

We provide a set of content dictionaries [SPSHP] for the formula translation. We jointly call them the SPSHP ontology. See Figure 2 for a depiction of the theory graph (a modular graph of theories that provide vocabularies of concepts and axiomatizations of the properties of their objects connected by theory morphisms – meaning-preserving transformations; see [Koh06; RK13]).

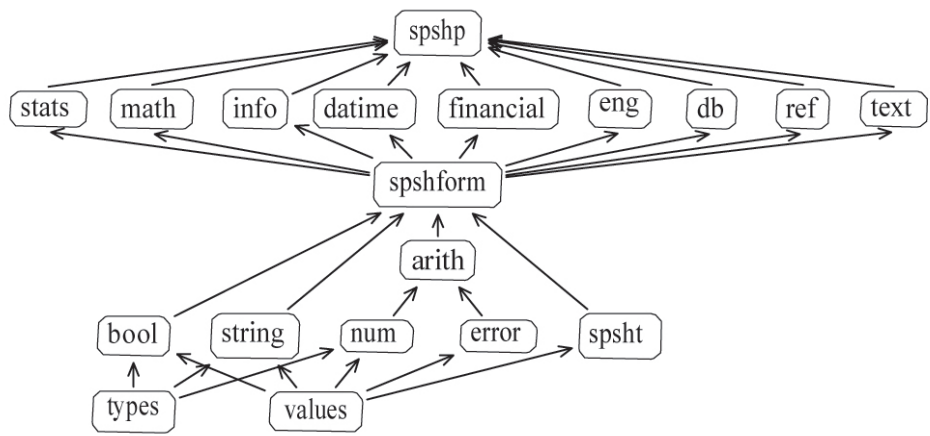

**Figure 2**: SPSHP: An Ontology for Spreadsheet Functions

**The SPSHP Content Dictionaries** The starting points of the SPSHP ontology in Figure 2 are the theories types (spreadsheets naturally induce a type system with flexary functions, optional arguments, and subtypes) and values which introduces the concept spreadsheet values. These are specialized into the subtypes for numbers (theory num with integers, floating point and complex numbers), strings (theory strings), and truth values (bool). The theory spsht provides the basic building blocks of spreadsheets (cells, rows, columns, tables) and their types. Theory error provides representations of typesheet errors raised by spreadsheet programs. Theory arith provides representations of the elementary arithmetic operations, which are not represented by spreadsheet functions but by the operators +, −, and ∗, etc. From all this material, theory spshform introduces the concepts of "value expressions" (expressions constructed from cell/range references, functions, strings, and numbers; they evaluate to spreadsheet values or errors) and value expression lists. Together with the flexary function types, the latter induce natural types of spreadsheet functions like SUM, which take arbitrarily many arguments that can be interpreted as lists of cell values. For instance, in the formula SUM(A5:A8,7,3) the range description A5:A8 induces a set of values in the spreadsheet computation. Theories stats to text provide the symbol declarations of the ca. 360 spreadsheet functions themselves; they follow the grouping found in spreadsheet applications. Finally, the theory spshp collects all the SPSHP sub-theories by importing them for convenience.

**Interoperability** So far, we have been able to keep the CDs in the SPSHP ontology independent of the particular spreadsheet application (MS Excel, OpenOffice Calc, Apple Numbers, Google Spreadsheet, etc.), as the formula languages of the applications have been standardized for interoperability. But there are functions whose implementations differ between applications, e.g. the COUNTIF function to count the number of cells which contain a certain value. If the cells A1 and A2 contain the value TRUE, then the formula COUNTIF(A1:A2;1) evaluates to 0 in Excel and to 2 in OpenOffice Calc.

In this case, we extend the SPSHP theory graph with application-specific theories as indicated in the picture on the right. Here, oo-stats and xls-stats are theories that specify functions whose semantics differ and that therefore cannot be specified in the application-independent theory stats. The theories ooc (for OpenOffice Calc) and xls (for MS Excel) are convenience theories, which collect the application- specific theories – note that by inheritance the theories *-stats contain all the functions from stats – just like spshp does in the application independent case. Intuitively, these theories represent the sub-ontologies for specific applications and are used for concrete translation projects. Note that e.g. ooc and xls share the majority of the specification and thus constitute a good basis for spreadsheet system interoperability (without translation) at the semantic level. However, even the application-specific functions are often aligned and very similar, thus we can specify views between the application-specific theories. OMDoc views[1] map concepts of the source theory to expressions in the target theory. For the view $j : \psi$, we have to implement the COUNTIF function from theory xls-stats in terms of the function COUNTIF from theory oo-stats, e.g. by removing truth values from the value formula lists in the arguments (the dual view $i : \Box$ can be defined similarly). Furthermore, the views between the application-specific component theories induce top-level views between theories ooc and xls that can (eventually) be used for semantic interoperation between spread- sheet applications, since they allow meaning-preserving translations of spreadsheet formulae.

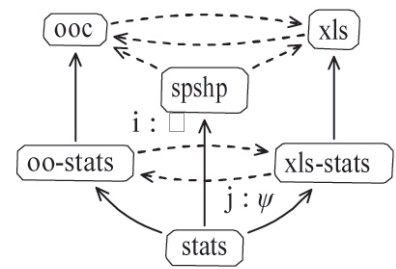

**Figure 3**: Interop. in SPSHP

---

[1] represented as dashed arrows in Figure 3; the label $j : \psi$ specifies the name i and the translation □.

3. **INDEXING AND QUERYING FORMULAE BY UNIFICATION**

MathWebSearch is an unification-based search engine for the efficient retrieval of mathematical formulae [MWS; KMP12]. The system consists of the three main components pictured in Figure 4. The *crawler subsystem* collects data from the corpora[2]. It transforms the mathematical formulae in the corpus into *MWS Harvest*s (XML files that contain formula-URIreference pairs) and feeds them into the core system. The *core system* (the MathWebSearch daemon mwsd) builds the search index and processes search queries: it accepts the MathWebSearch input formats (*MWS Harvest* and *MWS Query*; see [KP]) and generates the MathWebSearch output for- mat (*MWS Answer Set*). These are communicated through the *RESTful interface* restd which provides a public HTTP API conforming to the REST paradigm. The system supports two main workflows:

1. The crawler sends an *MWS Harvest* to mwsd. The XML is parsed and an internal representation is generated. This is used to update the Substitution Indexing Tree and consequently the database.
2. The user sends an *MWS Query* to mwsd. The XML is parsed, an internal query is generated. Using an efficient traversal of the index tree, formulas matching the search term are retrieved and aggregated into a result. This is translated to an *MWS Answer Set* and sent back to the user.

The system has been tested on large sets of formulae. Memory usage is linear (on average, 40 Mb for 1 Million formulae), while query times are fairly constant with respect to index size[3], averaging at 40 ms per query.

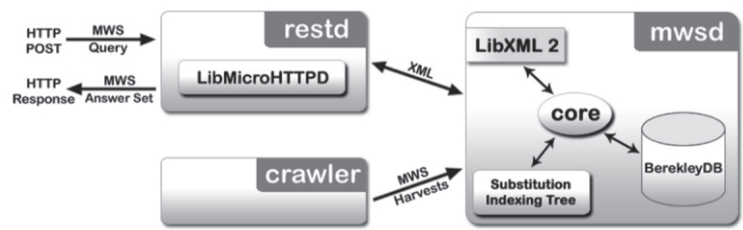

**Figure 4**: MWS-0.5 System Structure

---

[2]Note that we envision essentially one crawler per corpus. The crawlers are specialized to the respective formula representation, the organization and access methods to the corpus, etc.
[3]However, they do depend on the complexity of the query

(a) Spreadsheet with Linear Extrapolation with Legends    (b) Snippet of B7:F11

**Figure 5**: A spreadsheet and a cutout of a computed functional block

### 4. HARVESTING FORMULAE FROM SPREADSHEETS

In this section we describe the process of parsing spreadsheets and generating formula harvests that can be used by MathWebSearch. We are not only interested in the formulae, but also in the context they are used in. Therefore, we describe in Subsection 4.1 the context information we extract and the structure of the resulting harvest. The formula parser and converter is described in Subsection 4.2, the structure detection module for finding the context information is presented in Sub- section 4.3, while the harvest generator is described in Subsection 4.4. We describe the process of generating the harvest by using a slight modification of the Winograd spreadsheet from [KK09] (see Figure 5(a)). Our spreadsheet uses linear extrapolation for calculating the revenues and expenses in the projected years (see Section 1).

### 4.1 The Harvest Structure

Following [KK09], we use the term **legend** for those non-empty cells that do not contain input or computed values, but contain text strings that give auxiliary information on the cells that do. We call a grid region a **functional block** (FB), if that region could be interpreted as a function which maps elements from a legend to values. As the function is meant to be an intended function of the spreadsheet creator, it is immaterial whether the values are calculated or inputted. For example, the region `B13:F13` of Figure 5(a) could be interpreted as a function, which maps years to the total expenses in that year, and the region `B4:F4` as a function that maps a year to the revenues of that year. We call a functional block **computed** if all formulae are **cp-similar**, i.e. if they only differ in their cell references like `SUM(B4:B13)` and `SUM(C4:C13)`. Because all expenses for the projected years are calculated by linear extrapolation the area `E7:F11` in Figure 5(a) is a computed FB. A formal model which defines functional blocks and legends as mathematical objects is introduced in [Lig12].

To compute a harvest we need to find all computed functional blocks in a spreadsheet together with the parts of the legends surround them. For each computed FB, we create a harvest of the structure shown in Table 6. The contents of the surrounding legend cells are used as keywords which can be used to curtail the formula search. Because all formulae in a computed FB are cp-similar, we create one location-independent MathML representation per FB with the parser from Section 4.2. For representing a functional block as search result to a user, an XHTML snippet containing the FB and the surrounding legends is generated, like the one shown in Figure 5(b) for the FB `E7:F11`. Furthermore, the concrete formula of the upper-left cell from to FB is saved for search result representation. At last, the position information which locates the spreadsheet and the region in which the FB was found is also stored.

| MathML Formula |
|---|
| Position information |
| Keywords |
| Excel formula |
| XHTML Snippet |

**Table 6:** Harvest for an FB

### 4.2 Formula Parsing

We used the open source parser generator Antlr [Par13] to create a parser that transforms an Excel[4] formula into an abstract syntax tree (AST). Figure 7 shows the resulting AST for the formula `C7+(E$3-C$3)/(D$3-C$3)*(D7-C7)` from cell `E7`. The parser is aware of different operator priorities, nested formulae and cross worksheet references, and transforms cell references like `A5` to an integer based row and column pair. Creating MathML from ASTs is an easy programming exercise given a vocabulary of spreadsheet symbols that act as counterparts of the AST nodes. The SPSHP presented in Section 2 fills this requirement.

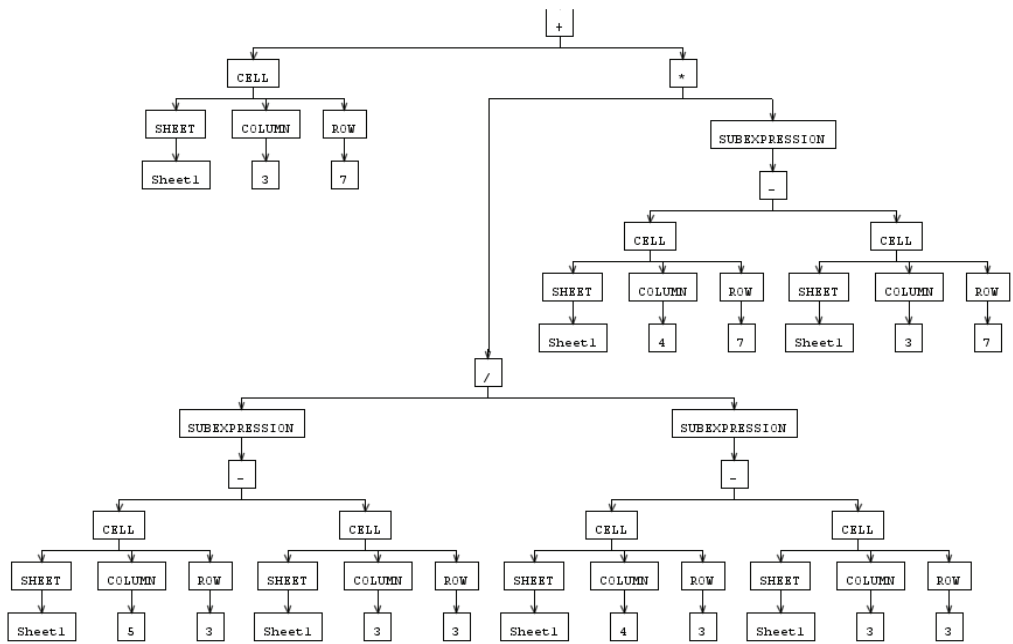

**Figure 7**: Abstract Syntax Tree of `C7+(E$3-C$3)/(D$3-C$3)*(D7-C7)`

### 4.3 Structure Detection in Spreadsheets

To find functional blocks and their legends, we use a simplification of our structure detection unit (SDU, see [Lig13]), which classifies each cell as "legend", "FB", "empty" or "hidden" and then aggregates regions into computed FB with legends.

---

[4] As formulae in other spreadsheet programs have nearly the same syntax as Excel, our parser can be tailored for those with minimal adjustments.

**Cell Classification** SDU uses a simple heuristics to classify some cells: formula cells are always "FB" and "nonempty", non-formula cells that contain at least 75% letters are classified as "legend". This heuristics are appropriate, because a misclassification of a non-formula functional block cell as legend is unproblematic, as it will just be integrated into the context of a computed FB (see "Area Detection" below). In particular, this heuristic correctly classifies the cells of Figure 5(a), except `B3:F3`, `B4:D4` and `B7:D11`[5]. Afterwards, hidden cells (like the cells `C1:F1`, `C2`, `D2` and `F2` in Figure 5(a)) are set to the type of the cell that hides the other ones (e.g. `C1:F1` are set to the type "legend" of cell B1 in Figure 5(a)).

**Area Detection** After classifying cells, SDU marks regions with cp-similar formulae as a functional block. In our example (see Figure 5(a)), we obtain the blocks `E4:F4`, `B13:F13`, `B15:F15`, and `E7:F11`, which SDU searches for the legends of each functional block. It starts in the first row of the FB and iterates upwards until it finds a row which contains at least one legend cell and no functional block cell in those cells that are right above the functional block. Then it iterates further upwards to the last row that is not empty and does not contain a functional block cell. The region between those rows which is right above the FB is taken as a legend region for the functional block. SDU repeats that search on the left side of the functional block and iterates through the columns instead of the rows. In our example in Figure 5(a) SDU finds a legend area in `E1:F3` and `A7:A11` for the functional block `E7:F11`.

### 4.4 Harvest Generation

For the generation of a XHTML snippet (see the one in Figure 5(b)) from the results of the area detection, we use the Apache POI API [POI] to get the relevant data from a spreadsheet. Therefore, we create a document representation of the original spreadsheet, and delete all worksheets except the one that contains the functional block. From the remaining sheet, we delete all rows and columns which do not contain a cell that is part of the functional block or surrounding legend. Afterwards, we use the HTML exporter from Apache POI to create an HTML document which is then transformed to XHTML by using JTidy [Jti].

For transforming a spreadsheet to a snippet, merged cells need some special attention. In our example, the cell `B1` in Figure 5(a) contains the header "Year" that is also relevant for the functional block `E7:F11`. Therefore, we move the content of merged cells that are partially inside and partially outside of a relevant legend region from the outside (e.g. from `B1`) to the inside part (e.g. to `E1`). As the HTML converter is not aware of merged regions, we delete all of them afterwards to avoid confusion.

### 5. XLSearch, A SEARCH ENGINE

We will now assemble a spreadsheet search engine from the components introduced above. Like any web search engine, XLSearch consists of a crawler, the core indexing/query engine (see Section 3), and a front-end that accepts queries and displays results.

**Crawler** As we imagine that the XLSearch engine will usually be deployed in institutional settings, which – after a spreadsheet audit – have created a spread-sheet inventory, we have restricted ourselves to a simple crawler that maps the MathML converter from Section 4.2 over a list of URIs of spreadsheets and generates MathWebSearch harvests from that are passed on to mwsd for indexing. But for the application in the search, we do not want concrete cell references in the index, since they are meaningless outside spreadsheet context. Therefore our parser variablizes cell and range references to MathWebSearch meta-variables (q:qvar in Figure 8), which can be instantiated in the search. In our example, the formula `C7+(E$3-C$3)/(D$3-C$3)*(D7-C7)` becomes the MathML expression in Figure 8.

---

[5]These cells can classified by other heuristics or via decision trees (see [Lig13]).

```
<math xmlns="http://www.w3.org/1998/Math/MathML"
xmlns:q=http://search.mathweb.org/ns>
  <apply>
     <csymbol cd="spsht-arith">opAdd</csymbol>
     <q:qvar name="X0"/>
    <apply>
       <csymbol cd="spsht-arith">opMul</csymbol>
      <apply>
         <csymbol cd="spsht-arith">opDiv</csymbol>
        <apply>
          <csymbol cd="spsht-arith">opSub</csymbol>
          <q:qvar name="X1"/>
          <q:qvar name="X2"/>
        </apply>
        <apply>
          <csymbol cd="spsht-arith">opSub</csymbol>
          <q:qvar name="X3"/>
          <q:qvar name="X2"/>
        </apply>
      </apply>
      <apply>
         <csymbol cd="spsht-arith">opSub</csymbol>
        <q:qvar name="X4"/>
        <q:qvar name="X0"/>
      </apply>
    </apply>
  </apply>
</math>
```

**Figure 8**: Index Entry for `C7+(E$3-C$3)/(D$3-C$3)*(D7-C7)`

**Front End** For simplicity, we use a web-based front-end that resembles web search engines for XLSearch; other front-ends, which e.g. embed XLSearch functionality into the spreadsheet program itself are imaginable, but are left to future research. Figure 9 shows a typical situation: the user has entered the query in the text box at the top. The query interface

1. accepts spreadsheet formulae in native syntax extended with query variables (names prefixed by ?)
2. converts them to MathML by the parser from Section 4.2 extended by a rule that transforms ?foo to <**q:qvar** name="foo"/>, and
    3. sends that to mwsd via its RESTful interface via a HTTP POST request.

In our example we see the formula ?fa+(?x−?a)/(?b−?a)*(?fb−?fa), which queries the index for linear extrapolation formulae.

mwsd returns a list of hits, all representing indexed formulae which unify with the query. Each hit carries a harvest datum as in Table 6 and keywords extracted from the containing FB, providing further information to the user. In Figure 9, the mwsd has found three hits. For each of these, the raw Excel formula, keywords and the URI reference (the URI of the spreadsheet and the FB identifier) are displayed. By clicking the second hit, a result snippet is revealed, in the form of the functional block with legends.

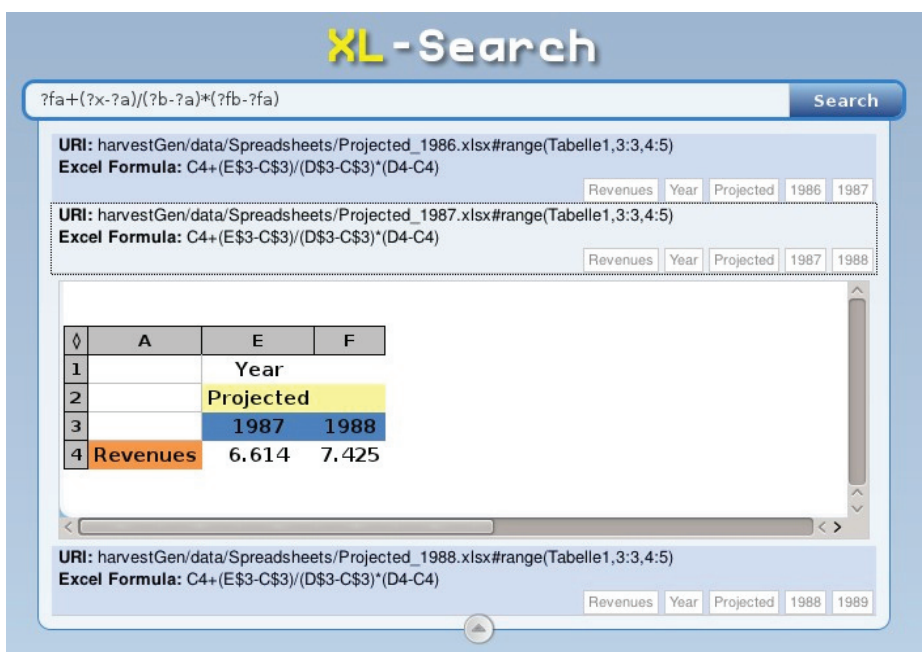

**Figure 9**: The XLSearch Web Front End

**Deployment & Demo** We have deployed an instance of XLSearch at http://search.mathweb.org/xl/ which indexes the EUSES corpus [FR05] with ca. 4.5 thousand spreadsheets.

6. **CONCLUSION**

We have presented a novel search engine that allows finding and accessing spreadsheets by their formulae. Such a search engine has multiple applications: it can be used to spot problematic formulae (e.g. known errors) in large spreadsheet corpora, or find re-usable tables (code blocks) in legacy spreadsheets leading to cost savings.

The main algorithmic core of the XLSearch engine is the pre-existing MathWebSearch formula search engine, which has been under constant development in our group for half a decade. For the application in the spreadsheet domain, we have de- veloped a standardized vocabulary (the SPSHP ontology) that allows to transform spreadsheet formulae into content MathML, which is the core of the input/query format of MathWebSearch.

**Further Applications** As the average query time is in range of 10-50 milliseconds, searches can even be utilized for very interactive applications. For instance, a variation of Netspeak [NSpk] for spreadsheet formulae. While Netspeak is able to find the most common word that is used in a phrasal context, our search engine finds the most common subformulae in a formula context. This can be very helpful for finding a very long and complex formula which can be just partially remembered by a user.

Alternatively, the spreadsheet system could monitor the number of similar formulae by sending off unification queries every time delimiters balance. As formulae in an organization are bound to be similar, an unexpected drop in the similar formula number could indicate a typo or error; and the author can be alerted in real time.

The SPSHP ontology supports applications in its own right: via the standardized format, formulae can be exported to other applications, e.g. via the clipboard (which supports MathML). Furthermore, formulae can be simplified or partially evaluated by standard symbolic computation systems, which can also also be used for query expansion, i.e., by searching for the variant `SUM(C7;(E$3-C$3)/(D$3-C$3)*(D7-C7))` of the linear interpolation formula.

**Future Work** Currently, the search engine hits are ranked by alphabetically sorting the file URIs. We expect that – as in Web search – ranking will be a crucial factor in the efficacy of search, and we want to explore this aspect further. We conjecture that for spreadsheets, where pagerank-like algorithms are hardly applicable, application-specific traits will have to be taken into account: [Sha+12] finds "*Studies suggest that location, file type, time, keywords, and associated events are the attributes best remembered*"; we are currently thinking about organizing search results by a file system tree widget with folding and unfolding interactions, if the corpus is organized this way.

Finally, we are thinking about including cognitive cues like the user-selected names for cells and ranges (see e.g. [Bew03; Spr]) into the search process as additional keywords.

**Acknowledgements** Work on the concepts presented here has been partially sup- ported by the German Research Foundation (DFG) under grant KO 2428/10-1 and HU 737/6-1 and the Leibniz association under grant SAW-2012-FIZ KA-2. The authors are indebted to the SiSsI group for discussions and insights on abstract spreadsheets.

## 7. REFERENCES

[Aus+10]  Ron Ausbrooks et al. *Mathematical Markup Language (MathML) Version 3.0*. W3C Recommendation. World Wide Web Consortium (W3C), 2010. http://www.w3.org/TR/MathML3.

[Bew03]  Philip L. Bewig. "In Excel, Cell Names Spell Speed, Safety". In: *Journal of Accountancy* (Nov. 2003). http://www.journalofaccountancy.com/issues/2003/nov/inexcelcellnamesspellspeedsafety.htm.

[Bur08]  Tim Burdick. *Improving Spreadsheet Audits in Six Steps*. 2008. url: http://www.theiia.org/intAuditor/itaudit/archives/2008/ march/improving-spreadsheet-audits-in-six-steps/.

[FR05]  Marc Fisher and Gregg Rothermel. "The EUSES Spreadsheet Corpus: A Shared Resource for Supporting Experimentation with Spreadsheet Dependability Mechanisms". In: *In 1st Workshop on End-User Software Engineering*. 2005, pp. 47–51.

[Jti]  *JTidy*. url: http://jtidy.sourceforge.net/ (visited on 04/08/2012).

[KK09]  Andrea Kohlhase and Michael Kohlhase. "Compensating the Computational Bias of Spreadsheets with MKM Techniques". In: *MKM/Calculemus Proceedings*. Ed. by Jacques Carette et al. LNAI 5625. Springer Verlag, July 2009, pp. 357–372. isbn: 978-3-642-02613-3. url: http://kwarc.info/kohlhase/papers/mkm09-sachs.pdf.

[KMP12]  Michael Kohlhase, Bogdan A. Matican, and Corneliu C. Prodescu. "Math- WebSearch 0.5 – Scaling an Open Formula Search Engine". In: *Intelligent Computer Mathematics*. Conferences on Intelligent Computer Mathematics (CICM) (Bremen, Germany, July 9–14, 2012). Ed. by Johan Jeuring et al. LNAI 7362. Berlin and Heidelberg: Springer Verlag, 2012, pp. 342–357. isbn: 978-3-642-31373-8. url: http://kwarc.info/kohlhase/papers/aisc12-mws.pdf.

[Koh06]  Michael Kohlhase. OMDoc – *An open markup format for mathematical documents [Version 1.2]*. LNAI 4180. Springer Verlag, Aug. 2006. url: http://omdoc.org/pubs/omdoc1.2.pdf.

[KP]  Michael Kohlhase and Corneliu Prodescu. *MathWebSearch Manual*. Web Manual. Jacobs University. url: https://svn.mathweb.org/repos/ mws/doc/manual/manual.pdf (visited on 04/07/2012).

[Lig12]  Christian Liguda. "Modeling the Structure of Spreadsheets". In: *Workshop on Knowledge and Experience Management*. Ed. by Kerstin Bach and Michael Meder. 2012, pp. 13 –17. url: http://dfki.de/~bach/FGWM-2012-Proc.pdf.

[Lig13]  Christian Liguda. "From Spreadhsheet Data to Structured Knowledge". manuscript, in preparation. Feb. 2013.

[MWS]  *Math Web Search*. url: https://trac.mathweb.org/MWS/ (visited on 01/08/2011).

[NO01]  David Nixon and Mike O'Hara. "Spreadsheet Auditing Software". In: *Symp. of the European Spreadsheet Risks Interest Group (EuSpRIG 2001)*. 2001.

[NSpk]  *Netspeak - One words word leads to another*. url: http://www.netspeak. org/ (visited on 04/17/2013).

[Pan00]  Raymond R. Panko. "Spreadsheet Errors: What We Know. What We Think We Can Do." In: *Symp. of the European Spreadsheet Risks Interest Group (EuSpRIG 2000)*. 2000.

[Par13]  Terence Parr. *The Definitive ANTLR 4 Reference*. Pragmatic Programmers, 2013.

[POI]  *Apache POI - the Java API for Microsoft Documents*. url: https:// poi.apache.org/ (visited on 04/08/2012).

[RK13]  Florian Rabe and Michael Kohlhase. "A Scalable Module System". Manuscript, submitted to Information & Computation. 2013. url: http://kwarc. info/frabe/Research/mmt.pdf. Submitted.

[Rot+01]  Gregg Rothermel et al. "A Methodology for Testing Spreadsheets". In: *ACM Transactions on Software Engineering and Methodology* 10 (2001), pp. 110–147.

[Sha+12]  Moushumi Sharmin et al. "On slide-based contextual cues for presentation reuse". In: *Proceedings of the 2012 ACM international conference on Intelligent User Interfaces*. IUI '12. Lisbon, Portugal: ACM, 2012, pp. 129–138. isbn: 978-1-4503-1048-2.

[Spr]  *Spreadsheet Page Excel Tips: Naming Techniques*. url: http://spreadsheetpage. com/index.php/tip/naming_techniques/ (visited on 04/08/2012).

[SPSHP]  *An Ontology for Spreadsheet Programs*. url: https://tnt.kwarc.info/repos/stc/projects/sissi/trunk/spshp